\documentstyle[epsfig]{article}
\newcommand{\institute}[1]{\parbox{16cm}{%
\centering\normalsize \sl #1}}
\vspace{2cm}

\def\lsi{\raise0.3ex\hbox{$<$\kern-0.75em\raise-1.1ex\hbox{$\sim$}}}
\def\gsi{\raise0.3ex\hbox{$>$\kern-0.75em\raise-1.1ex\hbox{$\sim$}}}

\title{A new method to study 
lattice QCD at finite temperature and chemical potential} 
\author{
Z.~Fodor$^{a,b}$ and S.D.~Katz$^b$\\
\institute{$^a$Deutsches Elektronen-Synchrotron DESY, Notkestr. 85, D-22607,
Hamburg, Germany\\
$^b$Institute for Theoretical Physics, E\"otv\"os University, P\'azm\'any
1, H-1117 Budapest, Hungary}}
\date{\today}

\begin{document}
\psfull

\maketitle

\begin{abstract} \noindent

Due to the sign problem, it is exponentially difficult to study QCD 
on the lattice at finite chemical potential. We 
propose a method --an overlap improving multi-parameter 
reweighting technique-- to alleviate this problem. We apply this
method and give the phase diagram 
of four-flavor QCD obtained on lattices $4^4$ and $4\cdot6^3$. 
Our results are based on ${\cal{O}}(10^3-10^4)$ configurations.
\end{abstract}

\section{Introduction}
Quantum Chromodynamics (QCD) at finite temperature ($T$) and/or 
chemical potential ($\mu$) is of fundamental importance,
since it describes relevant features of particle physics
in the early universe, in neutron stars and in heavy ion collisions.
According to the standard picture, as baryon density rises there is 
a change from a state dominated by hadrons --protons and neutrons--
to a state dominated by partons --quarks and gluons. In addition, 
recently a particularly interesting, rich phase structure has been 
conjectured for QCD at finite $T$ and 
$\mu$ \cite{qcd_phase,crit_point}. Of immediate interest is  
the existence and the location
of the critical point in the $T$-$\mu$ plane in three-flavor QCD, since
it can be explored by heavy-ion experiments. 
Clearly, the transition from hadronic to partonic (or to  
superconducting/superfluid) state is a fully 
non-perturbative phenomenon. 

Lattice gauge theory is a 
reliable systematic technique to study the non-perturbative 
features of QCD. QCD at finite $\mu$ can be 
formulated on the lattice \cite{HK83}; however, standard 
Monte-Carlo techniques can not be used at $\mu \neq 0$. The reason 
is that for non-vanishing real $\mu$ the functional measure 
--thus, the determinant of 
the Euclidean Dirac operator-- is complex. This fact
spoils any Monte-Carlo technique based on importance sampling. 

Several suggestions were studied in detail to solve the problem. 
We list a few of them.

In the large gauge coupling limit 
a monomer-dimer algorithm was used \cite{KM88}.
For small gauge coupling an attractive
approach is the ``Glasgow method'' \cite{glasgow} in which the 
partition function is expanded in powers of $\exp(\mu/T)$
by using an ensemble of configurations weighted by the $\mu$=0 action. 
After collecting more than 20 million configurations only unphysical
results were obtained: a premature onset transition. 
The reason is that the $\mu$=0 ensemble does not overlap sufficiently
with the finite density states of interest. 
Another possibility is to separate the absolute value and the
phase of the fermionic determinant and use the former to generate
configurations and the latter in observables \cite{T90}.

At imaginary $\mu$ the measure remains positive and standard Monte Carlo 
techniques apply. The grand canonical partition function can be obtained by
a Fourier transform \cite{imag,AKW99}. In this technique the dominant
source of errors is the Fourier transform rather than the poor overlap.
One can also use the fact that the partition function away from the 
transition line should be an analytic function of $\mu$, and the fit 
for imaginary $\mu$ values could be analytically continued to real 
values of $\mu$ \cite{L00}.  At T sufficiently above the transition, 
both real and imaginary $\mu$ can be studied by 
dimensionally reducing QCD \cite{HLP00}.
Hamiltonian formulation may also help studying the problem 
\cite{XQLUO}.
One can
study adjoint matter and color superconductivity in two-color QCD 
\cite{Hea00}. Nambu-Jona-Lasinio model
was also used as an effective theory for strong interactions 
\cite{NJL}. It has also been proposed to consider a $\mu$ 
for isospin rather than for baryon number \cite{AKW99,SS00}.
Another approach avoids the sign problem by using cluster algorithms,
in which negative against positive contributions are cancelled
\cite{CW99}. 

We
propose a method --an overlap improving multi-parameter
reweighting technique-- to reduce the overlap problem of the Glasgow method
and determine the
phase diagram in the $T$-$\mu$ plane. (Note, that a  
similar technique was successful for determining  
the phase diagram of the hot electroweak plasma \cite{ewpt}
e.g. on four-dimensional lattices, for which the applicability of a
single-parameter reweighting was poor.) 

We study the system (say four-dimensional QCD with dynamical fermions) 
at ${\rm Re}(\mu)$=0 around its 
transition point. Using a Glasgow-type
technique we calculate the determinants for each configuration for a
set of $\mu$, 
which, similarly to the Ferrenberg-Swendsen method \cite{FS89}, 
can be used for reweighting. Using 
the average plaquette values of the individual configurations for these 
partition functions an additional  
Ferrenberg-Swendsen \cite{FS89} reweighting can 
be done in the other parameter, thus in $\beta$, too. 
For Re($\mu$)$\neq$0 and/or Im($\beta$)$\neq$0 significant cancellations
of the complex phases appear, but exactly this feature is used 
in the determination of the zeros of the partition functions (Z), when
looking for the transition point.
Simultaneously reweighting in the two parameters $\mu$ and $\beta$
we can keep the system
on the transition line, which can be controlled e.g. by the inspection
of the Lee-Yang zeros \cite{LY52} of Z at complex $\beta$.
(Note, that the idea of performing a reweighting near the QCD critical
line was already suggested in \cite{AKW99}).
This technique gives a good overlap with the original 
transition-like states.
In principle any other parameter can be used for 
this type of reweighting.

We test this
method and illustrate its success compared to the Glasgow method. We 
present the exploratory results for the $\mu$-T phase diagram
of the dynamical $n_f$=4 staggered QCD. Simulations were done
on $L_t=4$ lattices of $4^4$ and $4\cdot6^3$. We estimate the phase 
diagram in physical units using the $\rho$ mass ($m_\rho$)
as the definition of the scale. 

Due to the small lattices and large 
spacings our estimate has systematic uncertainties, which
can be reduced by approaching the continuum limit. The study 
of this
limit is clearly not the goal of the present Letter. 

The Letter is organised as follows. Section II. presents  
the overlap ensuring multi-parameter reweighting technique. 
Section III. illustrates the applicability of the technique 
using $n_f$=4 dynamical QCD and gives the phase diagram in 
physical units. We conclude in Section IV.

\begin{figure}\begin{center}
\epsfig{file=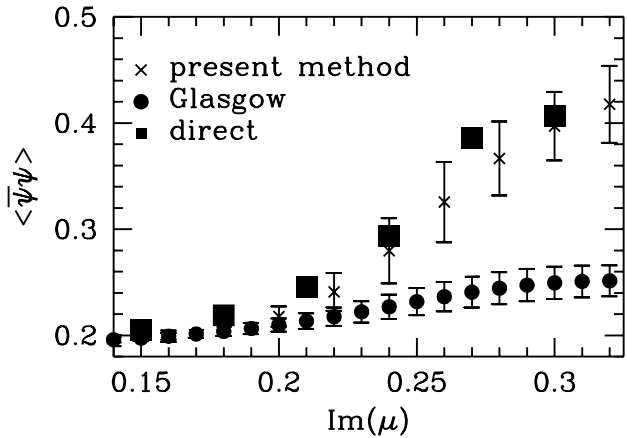,bb= 17 230 570 610, width=8.4cm}
\caption{\label{im_trans}
{The average of the quark condensates at $\beta$=5.085 
as a function of Im($\mu$), for
direct simulations (squares; their sizes give the errors), our technique (crosses)
and Glasgow-type reweighting (dots). 
}}
\end{center}\end{figure}

\section{Overlap improving multi-parameter reweighting}
Let us study a generic system of fermions $\psi$ and bosons $\phi$,
where the fermion Lagrange density is ${\bar \psi}M(\phi)\psi$.
Integrating over the Grassmann fields one gets:
\begin{equation} \label{part_func}
Z(\alpha)=\int{\cal D}\phi \exp[-S_{bos}(\alpha,\phi)]\det M(\phi,\alpha),
\end{equation}
where $\alpha$ denotes a set of parameters of 
the Lagrangian. Eg. in the case of QCD with staggered quarks $\alpha$ 
consists of $\beta$, 
the quark mass ($m_q$) and $\mu$ (which is included as 
$\exp(\mu)$ and $\exp(-\mu)$ multiplicative factors
of the forward and backward timelike links, respectively).
For some choice of the 
parameters $\alpha$=$\alpha_0$ 
importance sampling can be carried out (e.g. for Re($\mu$)=0). 
Rewriting eq. (\ref{part_func}) one obtains
\begin{eqnarray}
&&Z(\alpha)= 
\int {\cal D}\phi \exp[-S_{bos}(\alpha_0,\phi)]\det M(\phi,\alpha_0)
\nonumber\\
&&\left\{\exp[-S_{bos}(\alpha,\phi)+S_{bos}(\alpha_0,\phi)]
{\det M(\phi,\alpha)  \over \det M(\phi,\alpha_0)}\right\}.
\end{eqnarray}
Now we treat the terms in the curly bracket as an observable
--which is measured on each independent configuration-- 
and the rest as the measure. It is well known that changing 
only one parameter of the set $\alpha_0$ the ensemble 
generated at $\alpha_0$ provides an accurate value for some observables
only for very high statistics. This is ensured by important but 
rare fluctuations as the mismatched measure occasionally sampled the 
regions where the integrand is large. This is the so-called 
overlap problem. Note however, that we have several parameters
and the set $\alpha$ can be adjusted to $\alpha_0$ to ensure
much better overlap than obtained by varying only one, single 
parameter. Since the calculation of the determinants is 
expensive the most straightforward way to obtain a good overlap is to 
fix the parameters of the matrix M and
adjust the parameters which appear only in the bosonic action 
(in other words 
perform a Ferrenberg-Swendsen reweighting based on the bosonic part 
of the curly bracket). 

By simulating at a given set of bosonic couplings and using
Ferrenberg-Swendsen reweighting we can get information
on the system at other values of the couplings, 
even for complex ones. At finite volumes
Z($\alpha$) 
has zeros --thus the free energy singularities--
for complex values of these couplings. Standard finite size
scaling techniques can be used to analyse the volume dependence
of the Lee-Yang zeros. When looking for these  zeros 
we use reweighting for some bosonic couplings.  
Simultaneously changing two --or more--
parameters we can ensure that the system is reweighted along a 
transition line. This can be monitored by inspecting the
Lee-Yang zeros of $Z$.

\section{Illustration: $n_f$=4 dynamical QCD results}
For the case of lattice QCD at nonvanishing 
$\mu$ the above idea can be applied as
follows. We will use two parameter reweighting, namely 
reweighting in $\beta$ and $\mu$.
One performs the simulations at some $\beta,m$ and $\mu$ 
with ${\rm Re}(\mu)$=0 (note that purely imaginary $\mu$ can be directly
simulated). For each independent gauge configuration one 
calculates the average value of the plaquettes and 
the ratio of the determinants $\det M(\mu')/\det M(\mu;{\rm Re}(\mu)=0)$. 
For each $\mu'$ some $\delta \beta$ can be used
to reweight with the measured plaquette values. By this way a much
better overlap can be ensured than by reweighting only in $\mu$
(Glasgow method). 

We have tested these ideas in four-flavor  
QCD with $m_q$=0.05 dynamical staggered quarks. 
The molecular dynamics Monte-Carlo code of the
MILC collaboration was used \cite{milc}. 
We checked that the determinants
were calculated on independent configurations. 
Our statistical errors were obtained by a jackknife
analysis.  

\begin{figure}\begin{center}
\epsfig{file=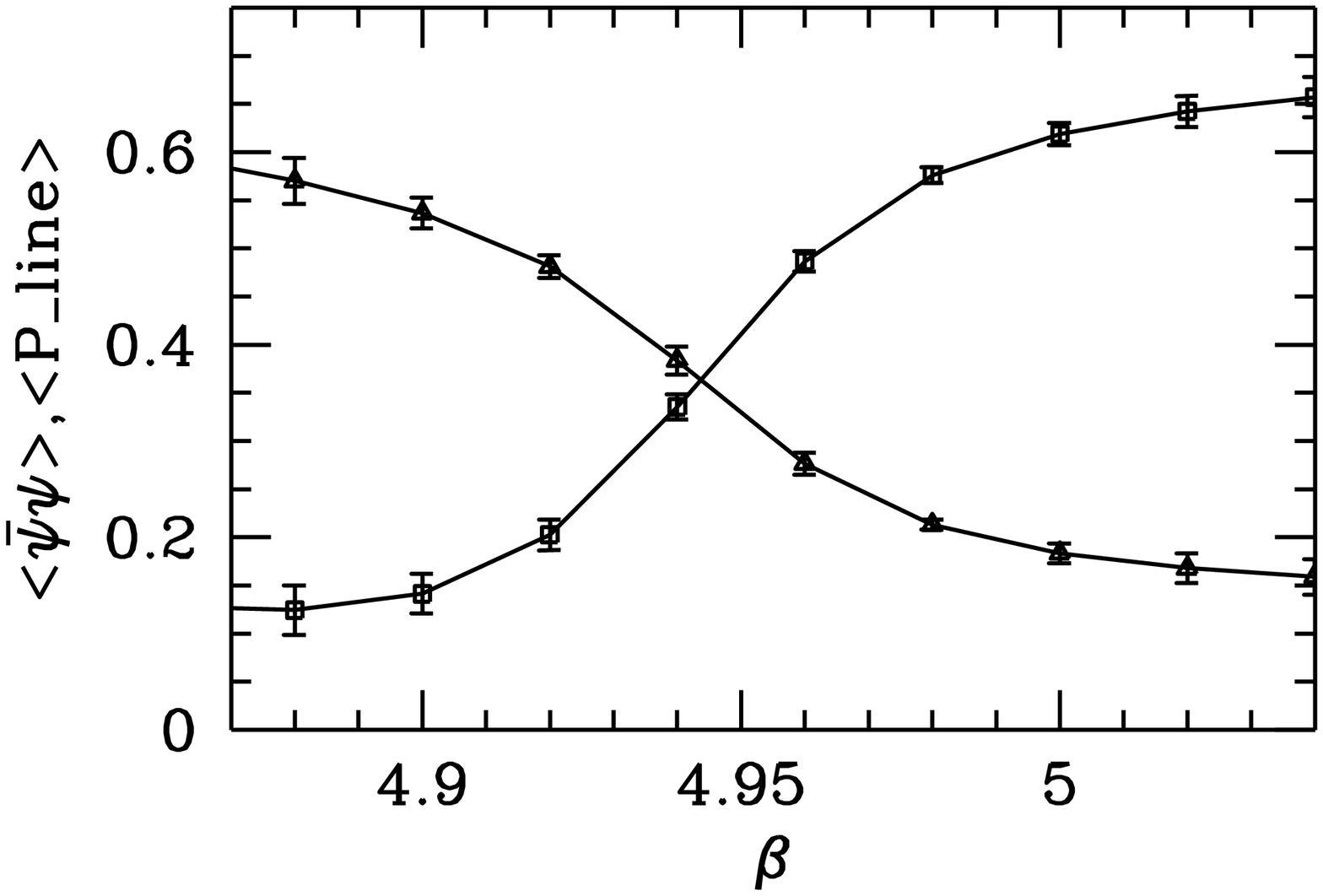,bb= 57 235 610 602, width=8.4cm}
\epsfig{file=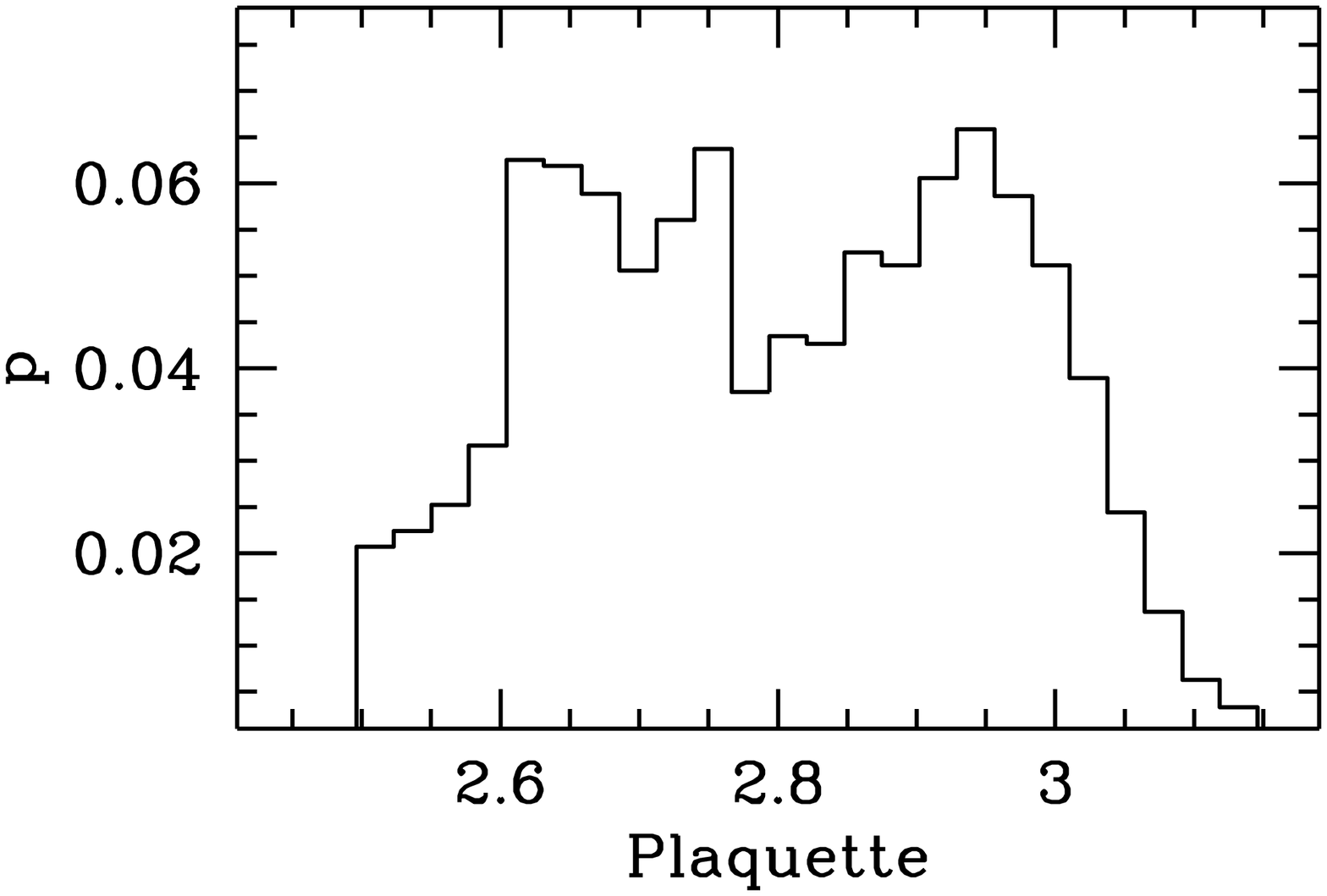, bb=-23 235 530 602, width=8.4cm}
\caption{\label{transition}
{The average of the Polyakov lines (squares) 
and quark condensates (triangles) as a function of $\beta$ at  
${\rm Im}(\mu)=0,\ {\rm Re}(\mu)=0.3$ (left panel). Histogram of
the plaquettes at $\beta$=4.938 and $\mu$=0.3 (right panel).
The lattice volume is $4^4$.}}
\end{center}\end{figure}

In order to directly check the applicability of our method we first 
collected 1200 independent V=4$\cdot 6^3$ configurations at 
Re($\mu$)=Im($\mu$)=0 and used the Glasgow-reweighting and 
also our technique to study Re($\mu$)=0, Im($\mu$)$\neq$0. 
For our method we used the transition $\beta$ (5.040) to
generate the configurations, while for the Glasgow method
$\beta$=5.085 was used. At 
Re($\mu$)=0, Im($\mu$)$\neq$0 and at $\beta$=5.085 
direct simulations are possible. 
After performing these direct simulations as well, a clear 
comparison can be done. Figure \ref{im_trans} shows the predictions of 
the three methods for the quark condensate.
The prediction of our method is in complete agreement with the direct 
results, whereas the predicition of the Glasgow-method is 
by several standard deviations off.
Figure \ref{im_trans} indicates that the      
two-parameter reweighting is far more trustworthy than the single 
parameter one. Note, that imaginary chemical potential is a useful check on
the proposed method, though it is different from a real chemical potential 
in biasing the ensemble towards non-zero baryon density. 

Based on these experiences we expect that our
method is superior also at
Re($\mu$)$\neq$0. 

Next we study the physical Re($\mu$)$\neq$0 case.
In order to have a better
overlap and to check further the applicability of our reweighting technique 
we carried out simulations on $4^4$ lattices at four different imaginary 
$\mu$ values ${\rm Im}(\mu)$=0,0.1,0.2,0.3. The results
obtained by the different runs are in complete agreement after
reweighting. In the 
following we use our largest sample generated at ${\rm Im}(\mu)$=0. 
On this smaller V
we used 13000 independent configurations.
On the $4\cdot 6^3$ lattice ${\rm Im}(\mu)$=0 was used with
1200 independent configurations. The runs were carried out at the transition
$\beta$ values at ${\rm Re}(\mu)$=0.

Let us illustrate that we are really at a transition point 
for nonvanishing $\mu$=0.3. Figure \ref{transition} shows the reweighted 
Polyakov-line and chiral condensate as a function of $\beta$. 
Furthermore, we check the correctness of our Lee-Yang reweighting approach
by showing the structure of a histogram. As it can be seen the turning point 
(indicating the coexistence of the two phases, thus the transition) is at 
$\approx \beta$=4.94. At $\mu$=0.3 the histogram method predicts
a equal-weight double-peak stucture at $\beta_c$=4.939(5), which is in
complete
agreement wit the prediction of the Lee-Yang zero technique (see later).

Table 1. gives the real parts
of the Lee-Yang zeros. Only zeros with the smallest imaginary parts
are listed. The real parts of these zeros are
usually used as a definition of the transition $\beta$ at finite V.
(Note, that the V$\rightarrow \infty$ limit of the imaginary 
parts tells the order of
the transition, cf. \cite{LY52,nf2}.)
Based on V=$4^4$ and $4\cdot 6^3$ 
we estimated the V$\rightarrow \infty$ limit by $1/V$ scaling. 
The critical $\beta$ in this limit 
is used to transform the results to physical units.  

\begin{table}[t]
\begin{center}\begin{tabular}{l|l|l}
${\rm Re}(\mu)$ & ${\rm Re}(\beta_0)$ for $V=4^4$ &
${\rm Re}(\beta_0)$ for $V=4\cdot 6^3$  \\
\hline
0.   &4.988(1) &5.040(2)  \\
0.05 &4.987(1) &5.038(2)  \\
0.1  &4.983(1) &5.033(2)  \\
0.15 &4.977(1) &5.023(2)  \\
0.2  &4.968(1) &5.009(2)  \\
0.25 &4.955(2) &4.993(3)  \\
0.3  &4.938(2) &4.975(3)  \\
0.35 &4.924(3) &4.965(4)  \\
0.4  &4.920(4) &4.959(4)  \\
\end{tabular}
\vspace{0.3cm}
\caption{\label{zeros}
{Lee-Yang zeros obtained at different $\mu$ values.
}}
\end{center}\end{table}

\begin{figure}\begin{center}
\epsfig{file=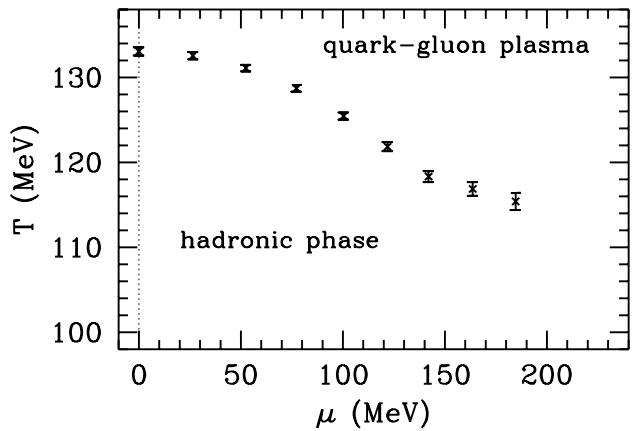,bb=17 230 570 602, width=8.4cm}
\caption{\label{diagram}
{The phase diagram in the $T$-$\mu$ plane for $n_f=4$ QCD. 
The physical scale is set by $m_\rho$. 
In physical units $m_q\approx$
25~MeV. The last point ($\mu \approx$190~MeV) corresponds to our largest
reweighted $\mu$.
}}
\end{center}\end{figure}

It is of particular interest to determine the phase diagram of
QCD on the $T$-$\mu$ plane.  
Though the lattices we used are absurdly small and the spacing is
quite large, it is  
illustrative to transform $\beta$, $m_q$ and $\mu$ into physical
units. Several parameters can be chosen to fix the scale, they
give quite different values at our $\beta$ couplings.
In the present analysis we fixed the scale by $m_\rho$=770 MeV. 
For small $\beta$ values, studied by the present paper, 
$m_\rho$ can be obtained by interpolating between the strong coupling
regime \cite{KS83} and the early measurements \cite{G88}. 
Figure \ref{diagram} shows the phase diagram
in physical units. 
The errorbars indicate the statistical 
uncertainties reached on our sample of only 
${\cal {O}}(10^3 - 10^4)$ configurations.
Note, that $L_t$=4 lattices with the above definition of the scale 
restrict  $T$ to be  larger than approximately 100 MeV (for
clarity we used this value at the origin).

\section{Conclusions, outlook}
We proposed a method --an overlap improving multi-parameter
reweighting technique-- to numerically study non-zero $\mu$ 
and determine the phase diagram in the $T$-$\mu$ plane. 
We applied this technique for $n_f$=4 QCD with dynamical staggered
quarks. 
We showed that for Im($\mu$)$\neq$0 the predictions of our method are 
in complete agreement with the direct simulations, whereas the Glasgow
method suffers from the well-known overlap problem.
Based on rather small statistics we were able to determine
the critical gauge couplings as a function of the real chemical potential,
which result was transformed into physical units. 

In this exploratory study we concentrated on the transition line
separating the two phases. Clearly, the same technique can be applied
to $T$-$\mu$ values somewhat below or above the line, for which the
configurations should be collected at an appropriately
chosen $\beta$ --below or above the transition one-- 
and at ${\rm Re}(\mu)$=0.

Note, that the factor in curly braces in (2) is of order $\exp[-V\delta]$,
with $\delta$, of course, dependent on the parameters $\alpha$ and $\alpha_0$
and the configuration $\phi$. Making $\delta$ small by a judicious choice
of $\alpha_0$ seems to work reasonably well at finite temperature.
Nevertheless, there are two  apparent limitations of the method. For large
enough volumes the reweighting factor will always be exponentially
suppressed, and at zero temperature $\delta$ is larger than at finite T, thus
the method most probably can not be applied to locate the transition at T=0.

Our method can be
easily applied to any number of Wilson fermions, whereas for 
$n_f$=2 or $n_f$=2+1 staggered fermions the situation is more
complicated due to the ambiguity of the roots of the determinants 
\cite{nf2}.

Note, that 
the present reweighting does not provide a general solution
to the sign problem in QCD, but could be useful to locate the critical
point of QCD, for which we are interested in a relatively small $\mu$ and
high T region of the phase diagram and may get away with relatively small
volumes. Another application is to determine the 
curvature of the phase diagram \cite{Eea01}.

We thank F. Csikor and 
I. Montvay for useful discussions and comments on the manuscript. 
This work was partially supported by Hungarian Science Foundation
grants No. 
OTKA-T37615/\-T34980/\-T29803/\-T22929/\-M28413-37071/\-OM-MU-708/\-IKTA111\-NIIF.
This work was in part based 
on the MILC collaboration's public lattice gauge theory code:
http://physics.indiana.edu/\~{ }sg/milc.html.

\end{document}